\documentclass[a4paper,fleqn]{cas-dc}

\usepackage[authoryear,longnamesfirst]{natbib}
\usepackage{lineno,hyperref}
\usepackage{ulem}
\usepackage{algorithm2e,setspace}
\usepackage{subfigure}
\usepackage{multirow}
\usepackage{bbding}
\usepackage{amssymb}

\def\tsc#1{\csdef{#1}{\textsc{\lowercase{#1}}\xspace}}
\tsc{WGM}
\tsc{QE}
\tsc{EP}
\tsc{PMS}
\tsc{BEC}
\tsc{DE}

\begin{document}
\let\WriteBookmarks\relax
\def\floatpagepagefraction{1}
\def\textpagefraction{.001}

\shorttitle{Underwater-art: Expanding information perspectives with text templates for underwater acoustic target recognition}

\shortauthors{Xie et~al.}

\title [mode = title]{Underwater-art: Expanding information perspectives with text templates for underwater acoustic target recognition}                      




%
\author[address1,address2]{Yuan Xie}[
                        style=chinese,
                        orcid=0000-0003-3803-0929]


\ead{xieyuan@hccl.ioa.cn.cn}



\credit{Conceptualization, Methodology, Software, Validation, Formal analysis, Investigation, Data Curation, Writing - Original Draft, Writing - Review \& Editing, Visualization}

\author[address1,address2]{Jiawei Ren}[style=chinese]
\ead{renjiawei@hccl.ioa.cn.cn}
\credit{Methodology, Investigation, Data Curation, Supervision}

\author[address1,address2]{Ji Xu}[style=chinese, orcid=0000-0002-3754-228X]
\ead{xuji@hccl.ioa.cn.cn}
\cormark[1]

\credit{Resources, Writing - Review \& Editing, Supervision, Project administration, Funding acquisition}

\affiliation[address1]{organization={Key Laboratory of Speech Acoustics and Content Understanding, Institute of Acoustics, Chinese Academy of Sciences},
    addressline={No.21, Beisihuan West Road, Haidian District},
    postcode={100190},
    city={Beijing},
    country={China}}
    
\affiliation[address2]{organization={University of Chinese Academy of Sciences},
    addressline={No.80, Zhongguancun East Road, Haidian District}, 
    postcode={100190},
    city={Beijing},
    country={China}}


\cortext[cor1]{Corresponding author}




\begin{abstract}
Underwater acoustic target recognition is an intractable task due to the complex acoustic source characteristics and sound propagation patterns. Limited by insufficient data and narrow information perspective, recognition models based on deep learning seem far from satisfactory in practical underwater scenarios. Although underwater acoustic signals are severely influenced by distance, channel depth, or other factors, annotations of relevant information are often non-uniform, incomplete, and hard to use. In our work, we propose to implement \textbf{U}nderwater \textbf{A}coustic \textbf{R}ecognition based on \textbf{T}emplates made up of rich relevant information (hereinafter called ``UART"). We design templates to integrate relevant information from different perspectives into descriptive natural language. UART adopts an audio-spectrogram-text tri-modal contrastive learning framework, which endows UART with the ability to guide the learning of acoustic representations by descriptive natural language. Our experiments reveal that UART has better recognition capability and generalization performance than traditional paradigms. Furthermore, the pre-trained UART model could provide superior prior knowledge for the recognition model in the scenario without any auxiliary annotation.
\end{abstract}




\maketitle

\section{Introduction}

Underwater acoustic target recognition is a vital component of marine acoustics. The task aims to realize automatic recognition based on the radiating sound of targets. Recognizing vessels from their sound could be meaningful for identifying the source of noise in underwater environmental monitoring systems~\citep{orr1994acoustic, fillinger2010towards, sutin2010stevens}.

In the last 20 years, the increasing demand has fostered research aiming at acoustic features and recognition algorithms. Mel scale-based spectrogram and Mel frequency cepstrum coefficients (MFCC), which utilize the mechanics of speech signals, were widely applied to underwater acoustic recognition tasks~\citep{lim2007classification, zhang2016feature, liu2021underwater, sun2022underwater}. Das et al.~\citep{das2013marine} used a cepstrum-based approach to realize marine vessel classification. Wang et al.~\citep{wang2014robust} used a bark-wavelet analysis combined with the Hilbert-Huang transform. Although designed in specific physics laws, low-dimensional features may inevitably cause the loss of information. Traditional machine learning methods could not achieve satisfactory accuracy on large-scale data with diversified features space~\citep{irfan2021deepship}. With the advent of deep learning, neural networks began to take the place of traditional recognition models. As reported in the literature, in the era of deep neural networks, researchers tended to use time-frequency spectrograms with comprehensive information as the feature. The Gabor spectrogram and wavelet spectrogram could localize the time-frequency domain information~\citep{david1990underwater, ou2009automatic, korany2016application, kubicek2020sonar}. LOFAR (Low Frequency Analysis Recording) could reflect the distribution of power spectrum and changes of signals in time and frequency dimensions~\citep{su2000multiple, wang2018underwater}. Such data-driven learning approaches tend to be competitive in complex underwater environments.

In practice, underwater acoustic signals are highly influenced by complex sound propagation patterns. The traditional paradigm prefers to label targets with categories while ignoring other influencing factors (e.g., distance of sound sources, channel depth, location, etc.). A question naturally arises: ``Could models comprehensively understand acoustic signals by just relying on the type of targets?” From the perspective of human perception, imparting a lot of knowledge based on a lack of overall perception is likely to distort our understanding of things. From a machine learning perspective, the type of targets alone is insufficient for models to comprehensively perceive the acoustic information, which may lead to undue inductive bias. Traditional paradigms only learn the mapping relation between audio and labels without considering any influencing factor. Affected by the changeable and unpredictable marine environment, the recognition performance of traditional models may be unsatisfactory. Considering the high correlation of underwater acoustic signals with environmental factors, the ``sciolism'' will result in severe overfitting and poor generalization performance in the actual situation.

In the underwater acoustic field, the influence of relevant factors (e.g., channel depth, location, distance, wind speed) on acoustic signals is non-ignorable. Although relevant factors usually contain rich environmental information, the annotations are often non-uniform, incomplete, and hard to use. Traditional frameworks seem to be incapable of utilizing such information. Inspired by the success of  CLIP-style~\citep{radford2021learning} contrastive learning on computer vision and natural language processing, changing the format of annotations from rigid labels to flexible natural language could realize unexpected good results. Some researchers have tried to transfer it to the audio field. There is some work on constructing a tri-modal contrastive learning model and applying it to the audio tasks~\citep{guzhov2021audioclip}~\citep{chen2022muser}. They proved that replacing labels with descriptive natural language that contains relevant information is novel and promising.

This paper proposes a recognition system - UART (Underwater Acoustic Recognition based on Templates). The core idea of UART is to take full advantage of relevant factors (e.g., channel depth, location, distance, wind speed) to guide the learning of acoustic representations. We apply manually designed text templates to integrate different forms of annotations into a unified text format (e.g., “The sound belongs to \uline{Fishboat}, which is in \uline{close} distance, and the channel depth is \uline{shallow}.”). Text templates not only integrate descriptive natural language sentences to provide broader perspectives, but also serve as decision boundaries for various task-related inferences. Moreover, we add an extra auxiliary encoder that takes spectrograms as input to broaden the model-level perspective and strengthen the robustness of the model. Thus, we realize the perspective broadening on both the data and the model level. We introduce the contrastive learning algorithm for training, and the flexible structure enables UART to accept non-uniform or incomplete data. In pursuit of practicality, we design a specialized inference strategy for UART. It only requires the audio to be predicted to accomplish inference.

We test UART on two underwater ship-radiated noise databases and implement several baseline methods on each dataset. UART shows better utilization efficiency of relevant information than baseline methods. We achieve the recognition accuracy of 92.74\% on Shipsear~\citep{santos2016shipsear} with the help of auxiliary annotations. Besides, the relevant information and the auxiliary encoder can bring a 6.6\% improvement in accuracy for UART, which is superior to our baseline approaches (2.59\% and 2.89\%). And for cases when there is only type information (no relevant information), pre-trained UART could also provide abundant prior knowledge. Using only 1\% data for training, pre-trained UART could achieve a 10.93\% performance advantage over traditional methods.

The main contribution of this paper is that we propose a novel framework – UART for underwater acoustic recognition. We innovatively use contrastive learning to replace the traditional feature-classifier paradigm. The advantages of UART are summarized as follows:

\begin{itemize}

\item UART uses text templates to integrate incomplete and non-uniform information into descriptive natural language. It solves the problem that auxiliary annotations are difficult to exploit, and reduces the requirements for data annotation quality.

\item UART applies contrast learning for training. We demonstrate that natural language-based text input could utilize auxiliary information more efficiently than discrete labels. UART could achieve 92.74\% accuracy on Shipsear.

\item UART flexibly supports simultaneous training of multiple encoders to broaden the perspective at the model level. Training with additional encoders such as spectrogram encoders proves to be beneficial.

\item The experimental results show that pre-trained UART could provide abundant prior knowledge regardless of auxiliary information, especially when the data resource is critically low.

\end{itemize}

\section{Framework and Methods}

\begin{figure*}
    \centering
    \includegraphics[width=0.75\linewidth]{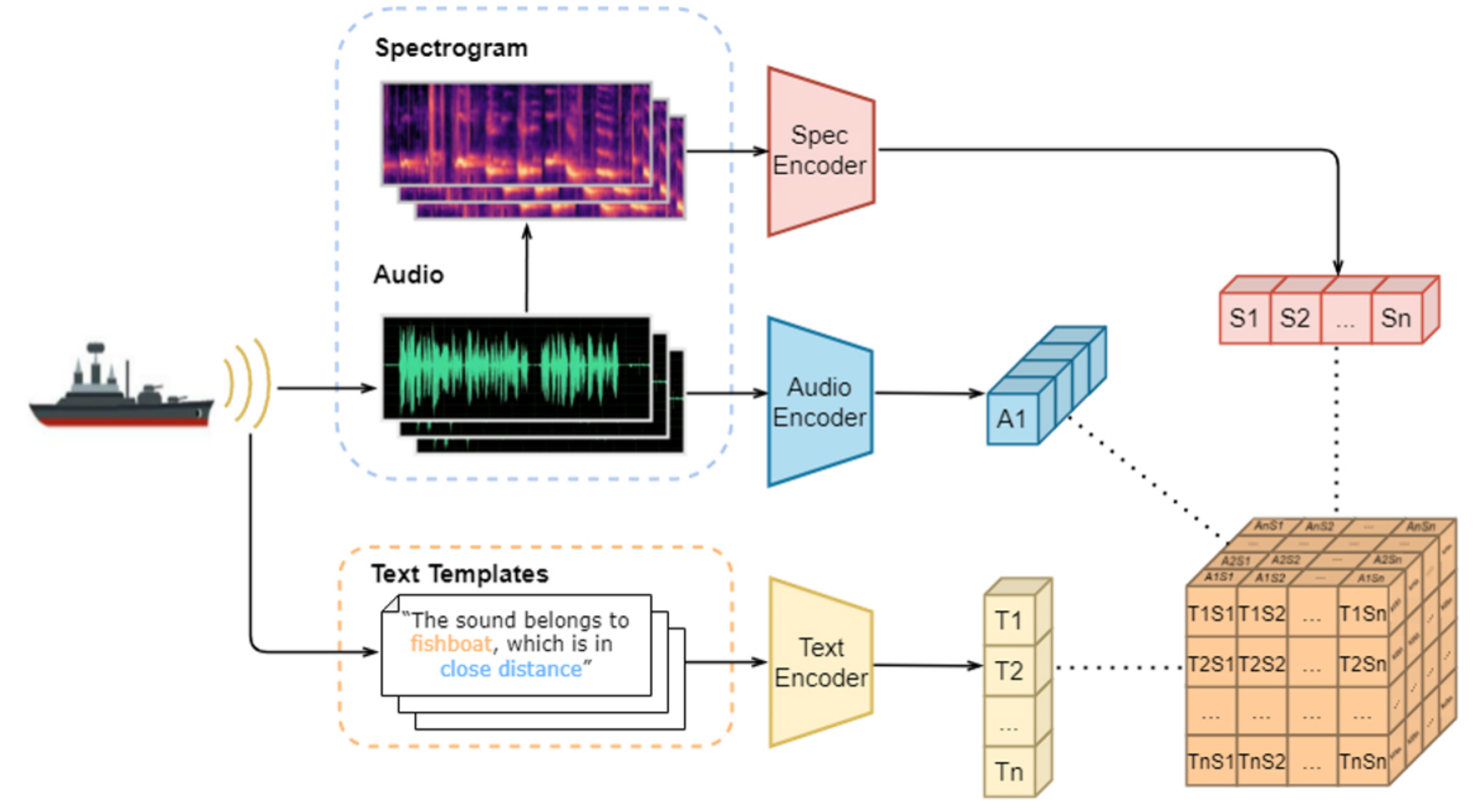}
    \caption{The framework of UART. UART consists of three encoders. The three encoders respectively map their inputs to embeddings of the same dimension. All embeddings are distributed in a shared embedding space.}
   \label{fig:framework}
\end{figure*}

\subsection{Data preprocessing}
Signals and corresponding annotations are available for the underwater acoustic recognition task. As illustrated in Fig~\ref{fig:framework}, UART needs to convert data into three modals: one-dimensional audio sequences, two-dimensional time-frequency spectrograms, and natural language sentences.

In this work, the input signal is the ship-radiated noise detected by the sonar array. In order to amplify signals in target directions and suppress interference and noise in other directions, array signal processing operations, such as beamforming, convert the data into mono audio signals.

Then, we calculate the time-frequency spectrograms based on the short-time Fourier transform (STFT) for signals. We perform STFT on signals by framing, windowing, and fast Fourier transform (FFT). After STFT, each frame will be spliced together according to the frequency axis to obtain the spectrogram. 

To obtain the text input, we make use of text templates to splice descriptive phrases or annotations into complete sentences. Text templates are designed in natural language form to facilitate the use of mature pre-trained language models. A simple template example is as follows: ``The sound belongs to $\lbrace$label$\rbrace$, which is in $\lbrace$distance$\rbrace$, and the channel depth is $\lbrace$depth$\rbrace$''.

\subsection{The Framework of UART}
As illustrated in Fig.\ref{fig:framework}, UART consists of three different encoders. The three encoders map the tri-modal inputs to a shared embedding space. The detailed introduction is as follows.

\textbf{Audio Encoder}: ESResNeXt~\citep{guzhov2021esresne} serves as our audio encoder. It is composed of two parts. The front end includes a learnable wavelet transformation with complex frequency B-spline wavelets~\citep{teolis1998computational}. Denote the input signal as $f(\cdot)$ and wavelet basis function as $\psi(\cdot)$. The translation parameter is $\tau$ while the scale parameter is $a$. The formula for wavelet transform is as follows:

\begin{equation}
    Wavelet(a,\tau) = \frac{1}{\sqrt{a}}\int_{-\infty} ^ {+\infty}f(t)\ast \psi(\frac{t-\tau}{a})dt.
\end{equation}

Complex frequency B-spline wavelet (Fbsp):
\begin{equation}
    \psi(x) = \sqrt{f_b}(sinc(\frac{f_b x}{m}))^m \exp(2i\pi f_c x)
\end{equation}

The adjustable parameters of the wavelet function contain order parameter $m$, bandwidth parameter $f_b$ and wavelet center frequency $f_c$. We enable $m$, $f_b$, and $f_c$ to participate in the gradient calculation process. Therefore, the values of parameters are automatically updated in a data-driven manner. The learnable wavelet transformation converts audio sequences into the wavelet spectrogram. After that, a ResNet-based~\citep{he2016deep} neural network receives wavelet spectrograms and outputs audio embeddings. Additionally, the network adds parallel convolution attention modules based on the ResNet backbone.

\textbf{Spectrogram Encoder}: We use a modified ResNet-50 as the backbone of the spectrogram encoder. We change the global average pooling layer into an attention pooling layer, where a ``transformer-style'' $QKV$ attention~\citep{wang2018non} is implemented. The spectrogram output from the encoder will serve as one of the embeddings.

\textbf{Text Encoder}: The text encoder is a Transformer-base~\citep{vaswani2017attention} network. Before operating on text tokens, the encoder employs a byte pair encoding (BPE) tokenizer to convert plain text into a sequence of discrete tokens. The text sequence will be bracketed with special tokens, \texttt{[SOS]} and  \texttt{[EOS]} . The activations of the \texttt{[EOS]} at the highest layer will be treated as the feature representation of the text and then linearly projected into the shared embedding space.

\subsection{Contrastive learning}
The core of UART is to utilize multi-modal information to optimize encoders simultaneously by contrastive learning. Given a batch of data, we mark the $i$th audio sample as $A_i$, and use pre-processing strategies to obtain the corresponding spectrogram input $S_i$ and the text input $T_i$. Then we get the embeddings of the three modals ($E_{A_i}$,$E_{S_i}$,$E_{T_i}$) by feeding inputs into their respective encoders. The optimization goal is to increase the similarity between embeddings from the same sample, while reducing the similarity between embeddings from different samples. Taking the data in the batch as an example, contrastive learning is to maximize the similarity between $E_{A_i}$ and $E_{S_i}$ ($E_{T_i}$) while minimizing the similarity between $E_{A_i}$ and $E_{S_j}$ ($E_{T_j}$) ($1\leq j\leq batchsize, j\neq i$).



From another perspective, the similarity represents the distance in the embedding space. What comparative learning does is to pull the $E_{T_i}$ closer to $E_{A_i}$ and $E_{S_i}$ while push the $E_{T_i}$ far away from $E_{A_j}$ and $E_{S_j}$ in the shared embedding space. To conclude, contrastive learning makes the similarity between embeddings more reasonable. In other words, it attempts to optimize the distance distribution in the embedding space. In this paper, we choose the cosine similarity function as the metric of similarity:

\begin{equation}
\begin{aligned}
    Similarity(E_{T},E_{S}) &= cos(\theta) = \frac{E_{T}\cdot E_{S}}{\Vert E_{T} \Vert_2 \times \Vert E_{S} \Vert_2} \\
    &=  
    \sum_{i=1}^n\frac{E_{T_i}}{\Vert E_{T} \Vert_2}\frac{E_{S_i}}{\Vert E_{S} \Vert_2}
\end{aligned}
\end{equation}

where $\theta$ refers to the angle between the embedding vectors. It is worth mentioning that we add an anomaly detection module during implementation. When the norm of embeddings is equal to 0, the model will automatically skip the corresponding input. It is not difficult to find that $\sum_{i=1}^nE_{T_i}E_{S_i}$ is similar to the form of diagonal matrix multiplication. For higher computational efficiency, we directly calculate the similarity matrix between modals. We denote the similarity matrix as $S$. The optimization goal is simplified to maximize the matrix element $S_{ii}$, while minimize the matrix element $S_{ij}$(j$\neq$i). It could be achieved by setting the loss function as the cross-entropy loss of the similarity matrix with the identity matrix. In addition, due to the huge difference in encoder structure, embeddings of different modals vary greatly in numerical values. Therefore, we need to normalize all embedding when computing cosine similarity. Besides, we add learnable scale coefficients to adaptively control the importance of different modals when calculating logits. A more detailed training algorithm is illustrated in Alg.\ref{alg:learning}.

\begin{algorithm}
\caption{Contrastive Learning of UART.}
\label{alg:learning}
\setstretch{0.9}
\SetAlgoLined

\KwData{Audio sequence (A), Text Template (T), Mel spectrogram (S), Ground Truth (Y)} 
(Note: $scale_{AT},scale_{TS},scale_{AS}$ represent learnable scale coefficients, Y is the identity matrix.)\\
\While{not done}{
    Sample batches $(A_i, T_i, S_i, Y_i)$.\\
    \ForAll{$(A_i, T_i, S_i, Y_i)$}{
        Compute embeddings $E_{A_i},E_{T_i},E_{S_i}$;\\
        as: e.g., $E_{A_i} = Encoder_{audio}(A_i)$\\
        Implement L2 normalization $\widehat{E}_{A_i},\widehat{E}_{T_i},\widehat{E}_{S_i}$; \\
        as: e.g., $\widehat{E}_{A_i} = \frac{{E}_{A_i}}{\Vert E_{A} \Vert_2}$\\
        Compute logits $Logits_{AT},Logits_{TS},Logits_{AS}$;\\
        as: e.g., $Logits_{AT} =Similarity_{AT}\cdot e^{scale_{AT}} =  \widehat{E}_{A_i}\cdot \widehat{E}_{T_i}\cdot e^{scale_{AT}} $\\
        Compute loss $\ell_{AT_i},\ell_{TA_i},\ell_{TS_i},\ell_{ST_i},\ell_{AS_i},\ell_{SA_i}$;\\
        as: e.g., $\ell_{AT_i} = CrossEntropy(Logits_{AT}, Y_i, axis=0)$\\
        \qquad \qquad   $\ell_{TA_i} = CrossEntropy(Logits_{AT}, Y_i, axis=1)$\\
    }
    Update encoders and weights with loss: $\mathcal{L} = (\ell_{AT_i}+\ell_{TA_i}+\ell_{TS_i}+\ell_{ST_i}+\ell_{AS_i}+\ell_{SA_i}) / 6$
}

\end{algorithm}

\subsection{Training and Tuning}

In this sub-section, we will introduce the training process of UART. In addition, we propose two application strategies for UART in different practical scenarios.

Alg.\ref{alg:learning} displays the training process of UART in detail. It initially accepts the raw mono audio sequences and descriptive annotations. After pre-processing, UART receives audio sequences, spectrograms, and text sentences as inputs. Three encoders convert their respective inputs into embeddings in a parallelized way. Three groups of embeddings share an embedding space. Then, according to Alg.\ref{alg:learning}, the loss could be calculated by contrastive learning, and UART will update the parameters of the three encoders according to the gradients simultaneously.

In order to make UART perform well in different data situations in practical scenarios, we propose two tuning strategies. First, we could pre-train UART on datasets with rich annotation information to obtain prior knowledge. Then we can decide which tuning strategy to use based on the data.

When the auxiliary annotations in practical tasks are abundant, the advantage of text input over discrete label input comes into play. The traditional paradigm needs to change the discrete label mapping and model structure. By virtue of the strong compatibility of text templates with different annotations, UART could adapt to new tasks just by modifying the text templates without changing the model structure and reconstructing the label map. The overall input form and training process keep the same as when training on practical tasks. We refer to the tuning process as ``UART-based tuning".

\begin{figure*}
    \centering
    \includegraphics[width=0.7\linewidth]{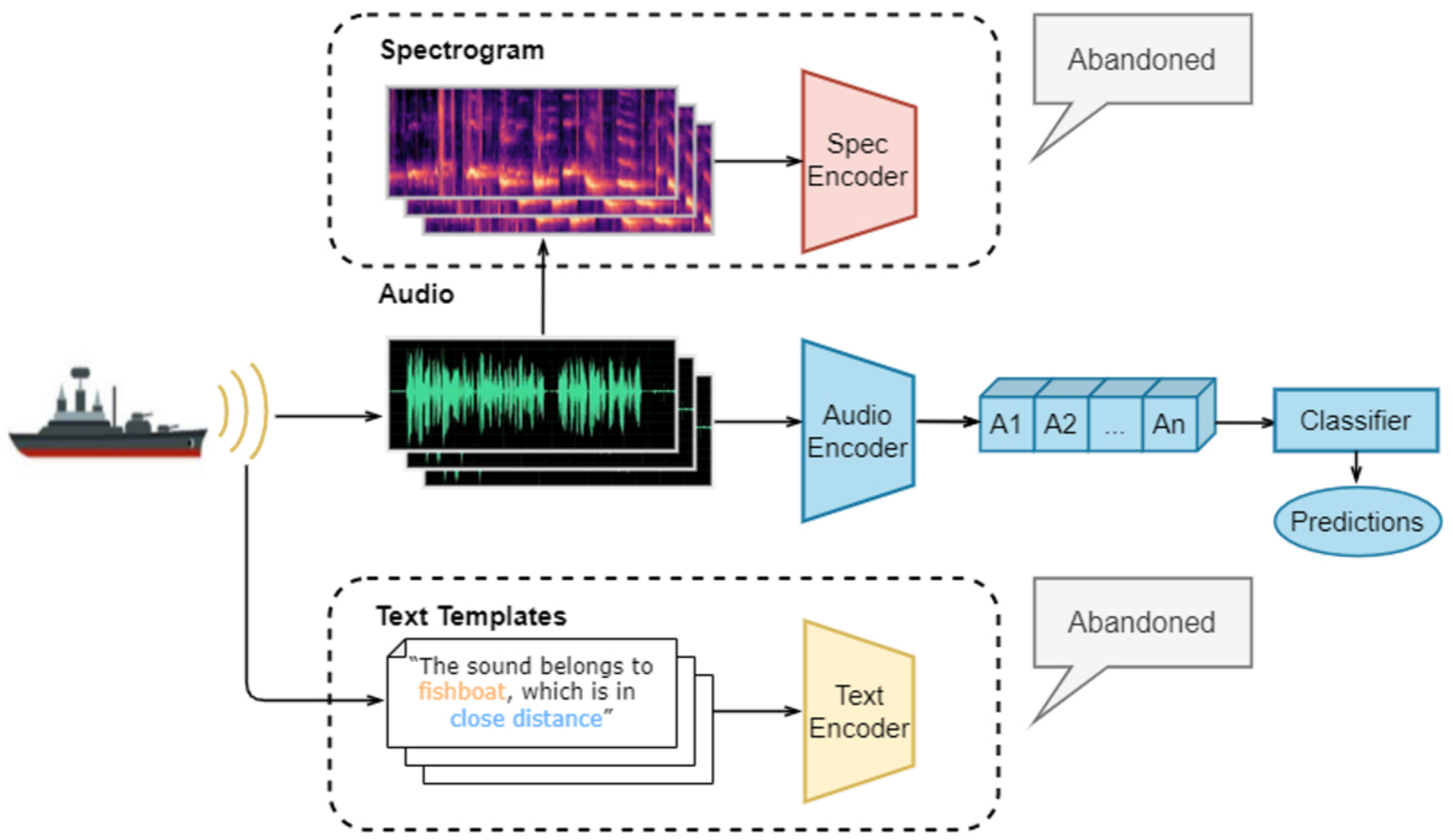}
    \caption{The process of encoder-based tuning. Encoder-based tuning only uses the audio encoder. The text encoder and spectrogram encoder are abandoned. A task-specific classifier needs to be added after the audio encoder.}
   \label{fig:Encode_tuning}
\end{figure*}


Another common situation in practical scenarios is that there are only category labels and no auxiliary information. UART-based tuning seems to be redundant in the absence of extra auxiliary information. In this case, we propose another tuning strategy. After pre-training our UART on the dataset with relevant information, we continue to train the audio encoder on practical tasks with only discrete labels. As depicted in Fig~\ref{fig:Encode_tuning}, we use discrete labels as input and abandon the spectrogram encoder and the text encoder because they seem redundant in the absence of auxiliary information. Then we add a task-related classifier to the bottom of the audio encoder for training. We refer to the tuning strategy as ``Encoder-based tuning”. At this time, the training paradigm changes from UART-based back to the traditional paradigm. ``Encoder-based tuning'' not only avoids the redundancy of the model structure, but also learns more prior knowledge from UART pre-training.

\subsection{Inference}

\begin{figure*}
    \centering
    \includegraphics[width=0.7\linewidth]{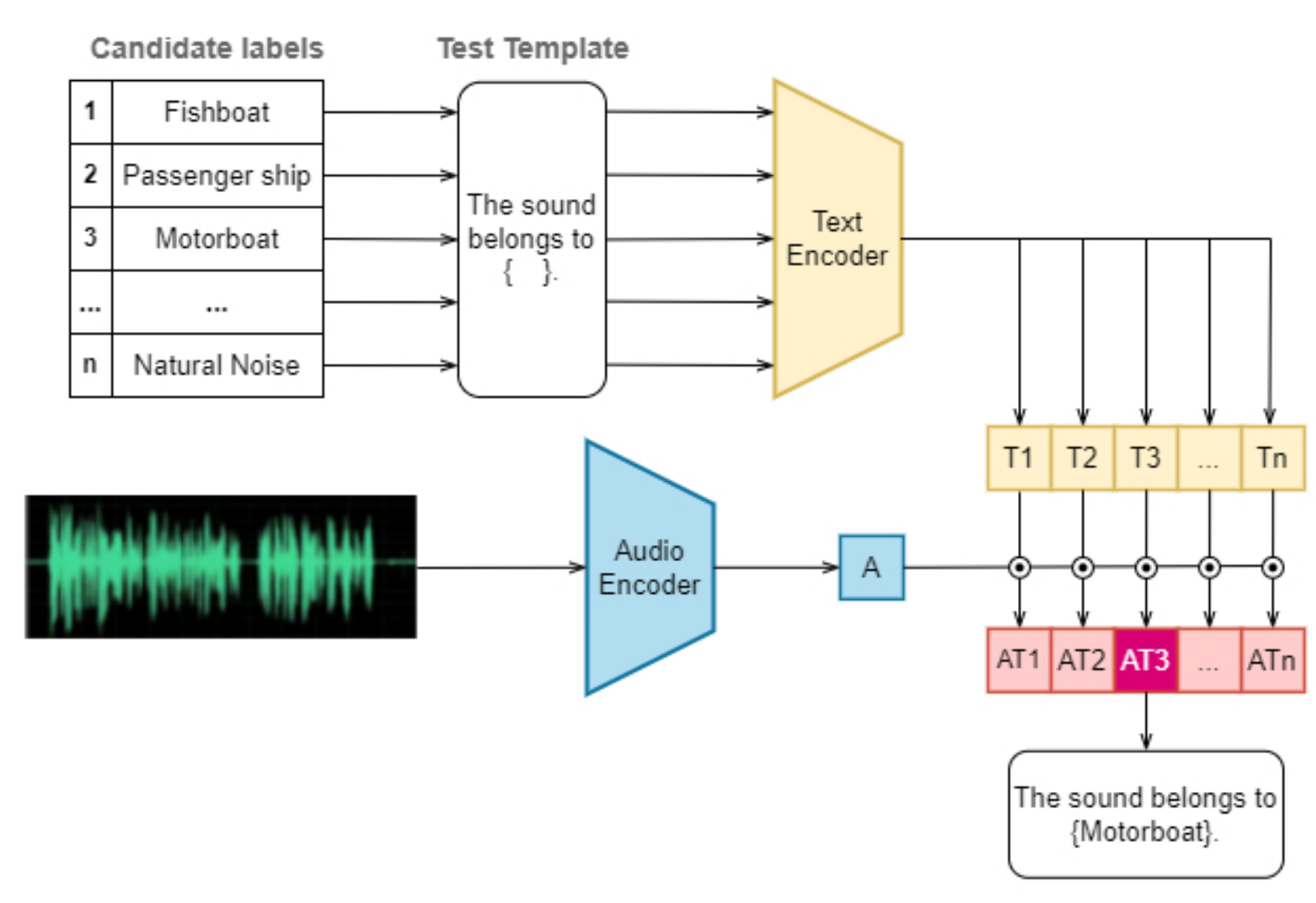}
    \caption{The process of inference. The label set to be predicted forms a candidate queue. All candidate labels are filled into the text template, then $N$ candidate embeddings are obtained via the text encoder. By calculating the similarity between audio embeddings and all candidate text embeddings, the one with the highest similarity is the prediction.}
   \label{fig:infer}
\end{figure*}

UART uses descriptive natural language to replace traditional labels, so it requires unconventional strategies for inference. Our strategy of inference is inspired by some work called ``Prompt''~\citep{radford2021learning, gao2020making}, we try to design a task-related test template (e.g., ``The sound belongs to $\lbrace$ $\rbrace$''), then fill all possible predictions into the test template to get a queue of candidates. As depicted in Fig~\ref{fig:infer}, we get the corresponding text embeddings by the text encoder. Then, we feed the audio sequence to be predicted into the audio encoder to get the audio embedding. Finally, we separately calculate the similarity between the audio embedding and all candidate text embeddings in the candidate queue. The label that corresponds to the text embedding with the highest similarity is our prediction. The entire process uses only audio to complete inference, it does not require any extra annotations.

For encoder-based tuning, the inference process keeps the same as traditional classifiers. The audio encoder converts the input audio sequences to audio embeddings, and the embeddings are converted into the predicted probability after the classifier and a softmax layer. We make predictions based on the output probability.


\section{Experiment Setup}

\subsection{Datasets}
Shipsear~\citep{santos2016shipsear} is a database of underwater recordings of ship and boat sounds. It is composed of 90 records (nearly 3 hours) representing sounds from 11 vessel types. It is hard to split little data into the form of ``train, validation, test'', so we select a subset of 9 categories (Dredger, Fishboat, Motorboat, Musselboat, Naturalnoise, Oceanliner, Passengers, RORO, Sailboat) for the recognition task. Besides label annotations, Shipsear also contains relevant information such as sound source distance, channel depth, location, and wind speed. Currently, most of the work on Shipsear divides all types into five classes by size. Table~\ref{tab:shipsear} shows the mapping. Although we use vessel types as the input to text templates during training, we calculate the accuracy according to the official classification criteria (Class A - Class E) to facilitate comparisons with other works on Shipsear.

\begin{table}[ht]
    \centering
	\caption{\label{tab:shipsear} Following the operations in ~\citep{santos2016shipsear}, the 9 vessel types (Dredger, Fishboat, Motorboat, Musselboat, Naturalnoise, Oceanliner, Passengers, RORO, Sailboat) were merged into 4 experimental classes (based on vessel size) and 1 background noise class.\\}
		\scalebox{1}{\begin{tabular}{cc}
		    \hline
		    Class  & Types\\
		    \hline
			A & Fishboat, Musselboat, Dredger\\
			B & Motorboat, Sailboat\\
			C & Passengers\\
			D & Oceanliner, RORO\\
			E & Naturalnoise\\
			
			\hline

         \\
		\end{tabular}}
\end{table}

DeepShip~\citep{irfan2021deepship} is an underwater acoustic benchmark dataset, which consists of 47 hours and 4 minutes of real-world underwater recordings of 265 different ships belonging to four classes (cargo, passenger ship, tanker, and tug). DeepShip only has category labels. There is no overlap between Deepship and Shipsear in terms of vessel types.  


\subsection{Unification and division of training data}
The two datasets used in this experiment have different sampling rates (52734Hz, 32000Hz). We downsample all audio files to 16000Hz. We cut each full audio into 30-second segments, and adjacent segments have an overlap of 15 seconds. To ensure that our results are credible and persuasive, we use 4-fold cross-validation for both Shipsear and DeepShip. In subsequent experiment results, the reported accuracy is the average results over four folds. Different segments of an audio sequence will be assigned to the same fold to ensure no overlap between the training set and the test set.  

\subsection{Baseline methods}
In order to make a fair comparison, we make sure that all comparison experiments are based on the same data and annotations.

\textbf{Multi-label~\citep{read2011classifier, tsoumakas2007multi}}: We fuse all available annotations to build a large label set. The new label set may contain categories, distance, location, or other annotations. Thus, one sample may be associated with multiple labels. We view the label set as a dictionary and unify the labels into the form of multi-hot. After the forward propagation and a SoftMax layer, we will get the predicted probabilities for each category. The category with the highest probability is our prediction.

\textbf{Multi-task~\citep{ruder2017overview}}: We create several tasks according to different types of annotations, such as category recognition, distance judgment, location recognition, or other tasks. We build a shared audio encoder and several task-related linear layers. Classifiers for different tasks consist of the shared encoder and corresponding task-related linear layers. The audio encoder could get trained for tasks simultaneously. After training, we accomplish inference using the shared audio encoder and the linear layer corresponding to the category recognition task.

\subsection{Parameters setup}
For framing, we set the frame length as 100ms and the frameshift as 50ms for all baselines. We implement the Mel-filter banks baseline by setting the number of filter banks to 300. For wavelet transform, we select the complex frequency B-spline wavelet as the wavelet basis function, and we follow the recommended parameter setting in the MathWorks wavelet toolbox ($m=2, f_b=0.5, f_c=1$). We also use this set of parameters as initial values of the learnable wavelet transform.

During training, we use the Adam optimizer~\citep{kingma2014adam} with weight decay regularization. We set a uniform learning rate to 1e-5 for all encoders and the weight decay to 1e-5. We train all models for 100 epochs. The entire UART model took 4 hours to complete training on Shipsear and 18 hours on DeepShip. All experiments are trained on 4 V100 GPUs.

\section{Results and discussion}
We first analyze some data in Shipsear to confirm the enormous impacts of factors (e.g., channel depth, distance, wind speed) on acoustic signals. Next, we select the best audio encoder for UART with satisfactory classification performance. Compared with several baseline methods, we test the results of UART on auxiliary annotations to prove the positive role of the text encoder and auxiliary information. Then we evaluate the effect of the spectrogram encoder by ablation experiments. Finally, we verify that the pre-trained UART could provide superior prior knowledge regardless of whether there is additional auxiliary information in practical scenarios.

\subsection{Data analysis}

\begin{figure*}
    \centering
    \includegraphics[width=0.75\linewidth]{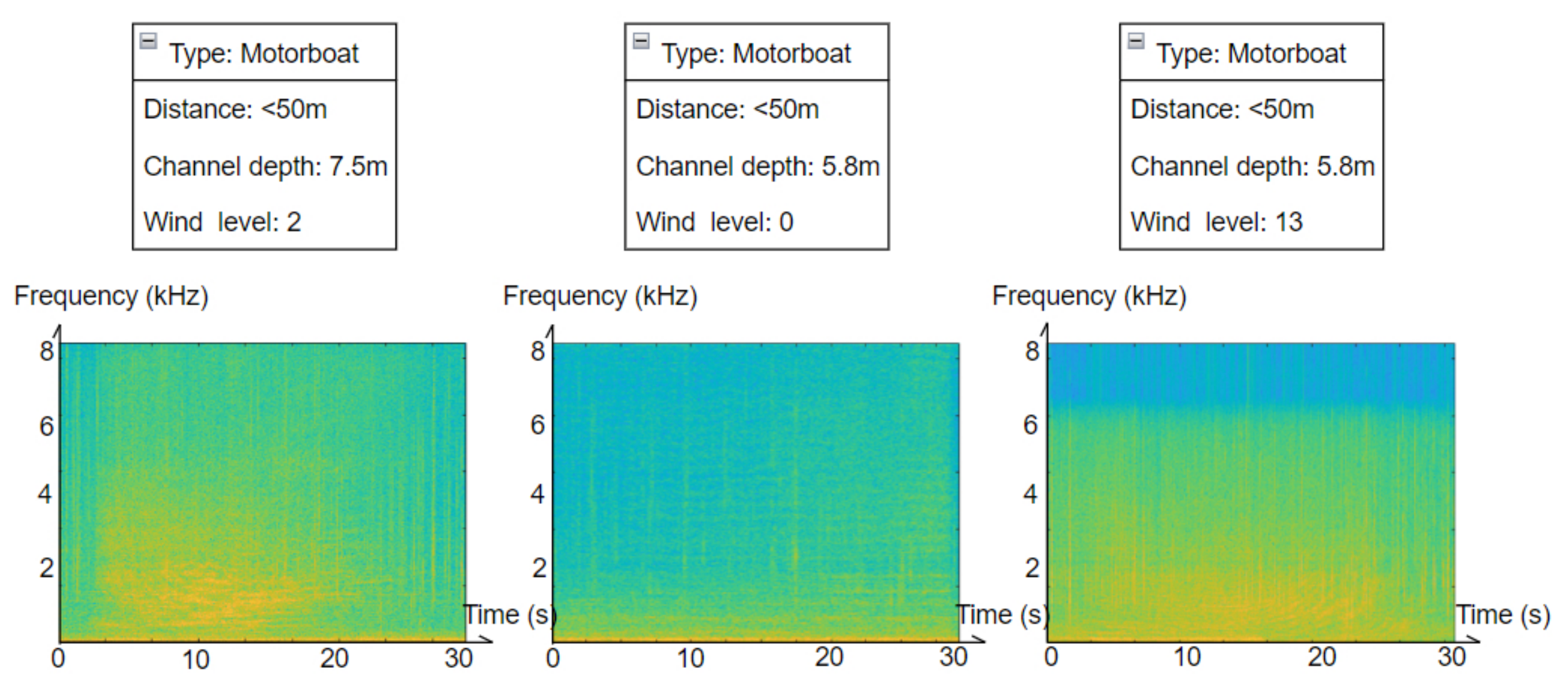}
    \caption{Time-frequency spectrograms of three samples belonging to the type of motorboat. Their respective distance, channel depth, and wind level information are marked in the figure.}
   \label{fig:data_analysis}
\end{figure*}

As the starting point for our research, underwater acoustic signals are severely disturbed by various factors. To ensure the rigor of our work, we analyze some data in Shipsear and visualize several samples using time-frequency spectrograms. As shown in Fig~\ref{fig:data_analysis}, we plot the time-frequency spectrograms for motorboats with different environmental conditions. In this set of comparisons, there exist huge differences between the spectrograms. It seems not easy to recognize them as the same type (motorboat). The influence of environmental conditions on underwater acoustic signals is drastic, and multivariate analysis is very complex. So we propose UART to comprehensively model all relevant information.

\subsection{Selection of the Audio Encoder}
The main structure of UART is composed of three encoders. Therefore, the choice of the encoder structure is crucial. We select ResNet-50 for the spectrogram encoder and Transformer-base for the text encoder, which has been widely used and proved reliable. However, the choice of the audio encoder is controversial. We implement several baseline models that could be used as encoders and compare their performance. Results in Table~\ref{tab:table1} show that the learnable wavelet spectrogram may be our best choice. It could achieve a recognition accuracy of 86.15\% on Shipsear by only labels of vessel types. Its learnability enables it to get more benefits with the help of natural language supervision. In all following experiments, we uniformly used learnable wavelet spectrogram \& ResNet50 as the structure of UART's audio encoder.

\begin{table}[ht]
    \centering
	\caption{\label{tab:table1} Comparisons of different audio encoders on Shipsear. Time-domain features include energy, zero-crossing wavelength, zero-crossing wavelength difference, peak-to-peak amplitude, etc. All results are measured in 30 seconds. We take the average accuracy on 4 folds as the metric.}
	\scalebox{0.65}{
		\begin{tabular}{lcc}
            \hline
			Feature/Input& Model  & Accuracy\\
			\hline
			Time domain features& ResNet50 &  71.90\\
			DEMON~\citep{lu2020fundamental} & Linear & 78.79\\
			Mel-Filter Banks~\citep{zhang2016feature}& ResNet50 & 84.90\\
			STFT spectrogram & ResNet50 & 84.89\\
			Gabor spectrogram~\citep{shastri2013time}& ResNet50 & 85.15\\
			Wavelet spectrogram~\citep{courmontagne2012time}& ResNet50 & 85.24\\
			Learnable wavelet spectrogram~\citep{guzhov2021esresne} &ResNet50 & \textbf{86.15}\\
			\hline
		\end{tabular}}
\end{table}

\subsection{Role of text templates}

\begin{table}[ht]
	\caption{\label{tab:table2} Comparisons of the performance of UART with different auxiliary information for training. The experiment is carried out on Shipsear. The audio encoder keeps the same structure for all models and all methods. All results are measured in 30 seconds. We take the average accuracy on 4 folds as the metric.}
		\scalebox{0.75}{\begin{tabular}{lcccc}
            \hline
			Auxiliary information &Single-label &Multi-label& Multi-task & UART\\
			\hline
			- &86.15&-&- & 84.26\\
			Distance &-&86.11&87.33&90.12\\
			Depth &-&87.39&87.94&89.62\\
			Local &-&86.94&89.04&89.65\\
            Distance,Depth&-&87.09&86.00&\textbf{90.69}\\
			Distance,Local&-&88.67&86.17&90.27\\
			Depth,Local&-&88.24&87.38&88.74\\
			Distance,Depth,Local&-&87.87&85.05&89.51\\
			\hline
         \\
		\end{tabular}}
\end{table}

\begin{figure*}
    \centering
    \includegraphics[width=0.75\linewidth]{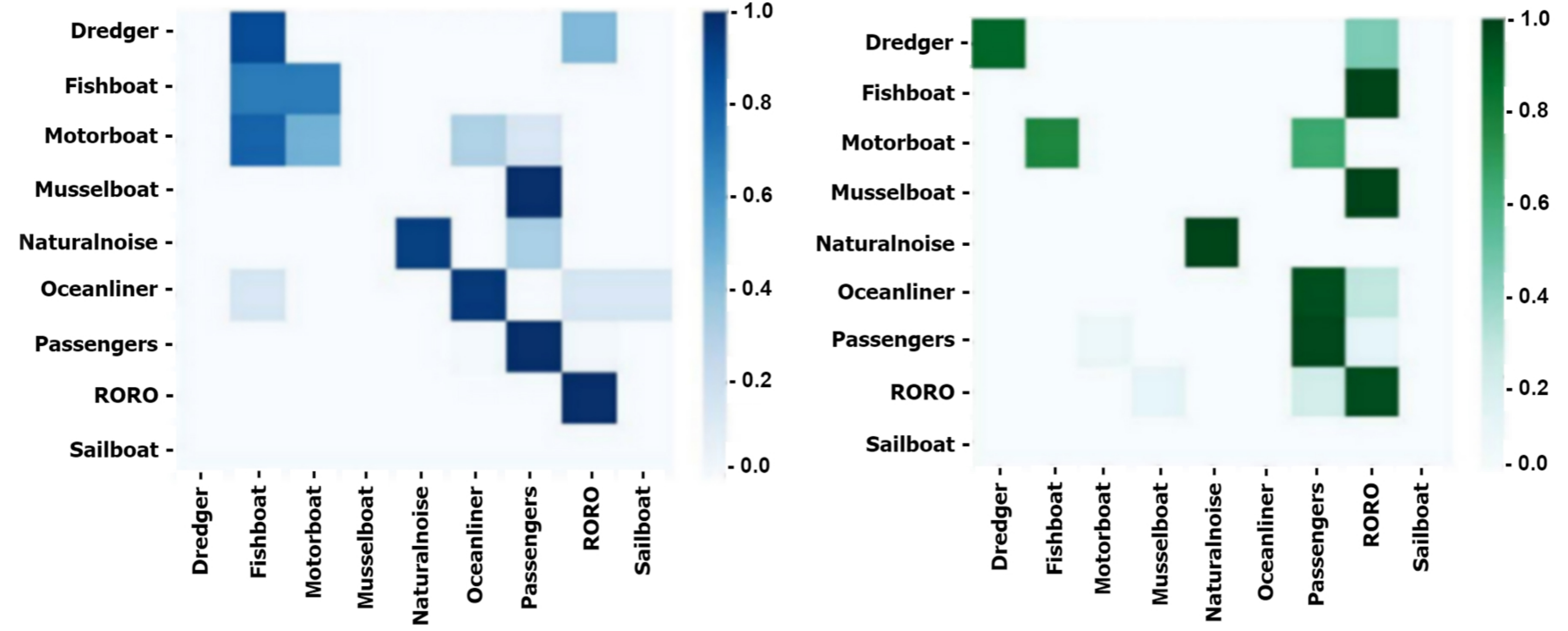}
    \caption{The confusion matrix for classification on Shipsear (Dredger, Fishboat, Motorboat, Musselboat, Naturalnoise, Oceanliner, Passengers, RORO). The former represents the classification results using auxiliary information, while the latter represents the results using only labels. The depth of the color represents probability.}
   \label{fig:confusion}
\end{figure*}

In this sub-section, all experiments select UART as the model. We get different text inputs by changing auxiliary information for comparison. In Table~\ref{tab:table2}, in the case of training with only labels of vessel types, we observe that the performance of UART is slightly inferior to that of traditional training methods using discrete labels. With the addition of auxiliary information, different training strategies all achieve some degree of performance improvement, while UART begins to show its promising potential and gets a significant performance boost. The multi-label baseline gets a 2.52\% boost and the multi-task baseline gets a 2.89\% benefit, while the accuracy rate of UART arises from  84.26\% to 90.69\%. It reveals that UART benefits most and outperforms all baseline methods. From the confusion matrix in Fig~\ref{fig:confusion}, we find that some classes that were completely unrecognizable (e.g., Oceanliner, Fishboat) could be recognized correctly after using auxiliary information. It shows that some of their latent characteristics are dug out by text inputs with abundant semantic information.

By Analyzing the data in Table~\ref{tab:table2}, the multi-label method gains little from extra information. It mixes all information and feeds it into the model without any boundary. The model swallows this knowledge ponderously and has no capability to understand the inherent meaning. It may be why Multi-label gains little benefit from extra information. For the multi-task model, different tasks will interact mutually, while the influences are not always positive. Tasks with low relevance or too much information can easily skew the model weights. In other words, the auxiliary task may bring the model weight to the local optimal point, resulting in only sub-optimal solutions for the main task. It leads to large fluctuations in the multi-task mode. The accuracy fluctuations in Table~\ref{tab:table2} prove this point.

It is not difficult to see that UART makes efficient use of additional information. When only using labels, the operation of converting discrete labels to text templates seems to be redundant. The boundary of templates is superfluous in this situation. With the addition of auxiliary information, text templates begin to realize their promising potential. UART performs better than baseline methods regardless of the content and form of the extra information at all times. Models based on labeling data are unaware of semantic information and the relationships between labels. It makes the decision boundary tend to be random in the embedding space. For instance, ``windy'' is similar to ``breeze'', but not similar to ``shallow water''. Discrete labels will only map them to different discrete numbers. If not well initialized, the distance distribution in the embedding space may be unreasonable (e.g., embeddings corresponding to ``windy'' and embeddings corresponding to ``breeze'' are far away). For UART, the pre-trained Transformer encoder could extract the semantic information of the text sentences, so as to ensure that the distance distribution of various annotations in the embedding space is reasonable (e.g., embeddings corresponding to ``... the weather is windy'' must be close to embeddings of ``... the wind is breeze''). It enables UART to learn correct and valid knowledge rather than being overwhelmed by complex auxiliary information.


\subsection{Analysis of auxiliary information}

\begin{figure*}
    \centering
    \subfigure[Distance distribution of different classes of ship noise. The vertical axis shows the proportion of the class.]{
        \begin{minipage}[b]{1\textwidth}
        \includegraphics[width=0.75\linewidth]{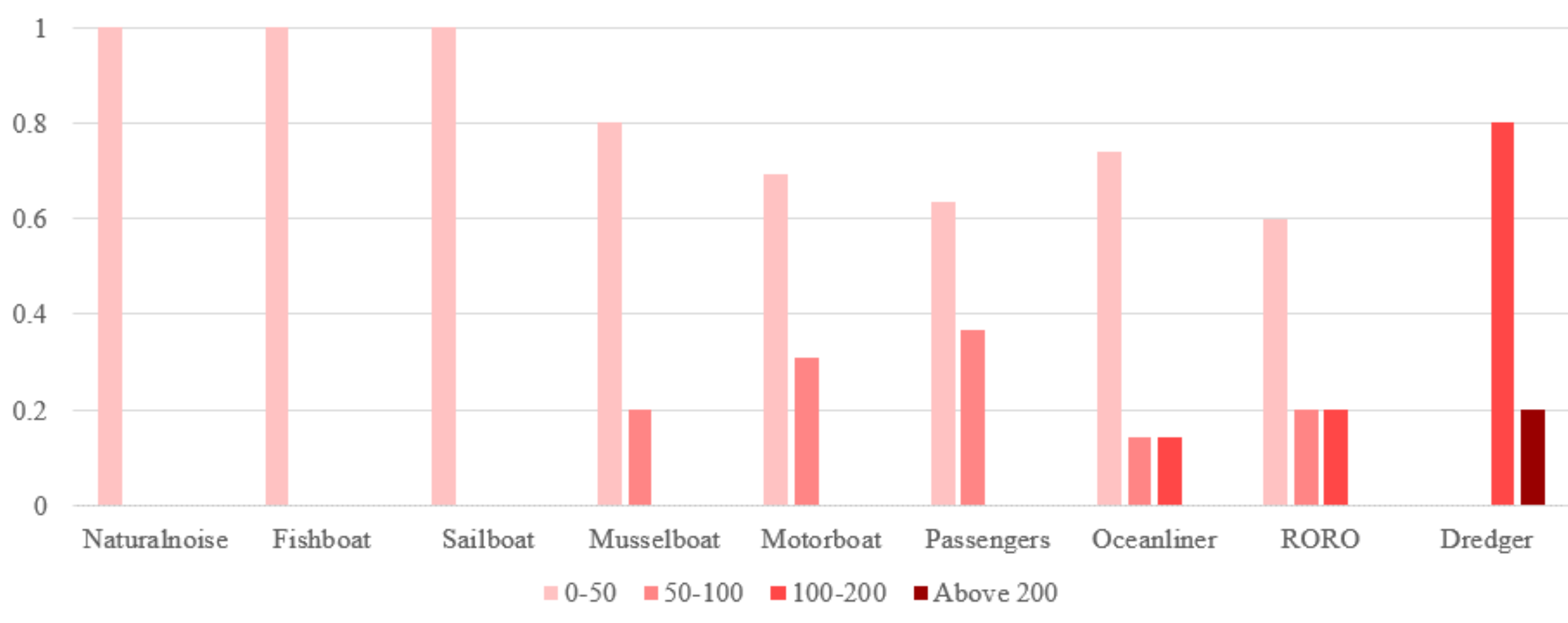}
        \centering
        \end{minipage}}

    \subfigure[For RORO ships, time-domain envelope diagrams for different sound source distances.]{
        \begin{minipage}[b]{1\textwidth}
        \includegraphics[width=0.75\linewidth]{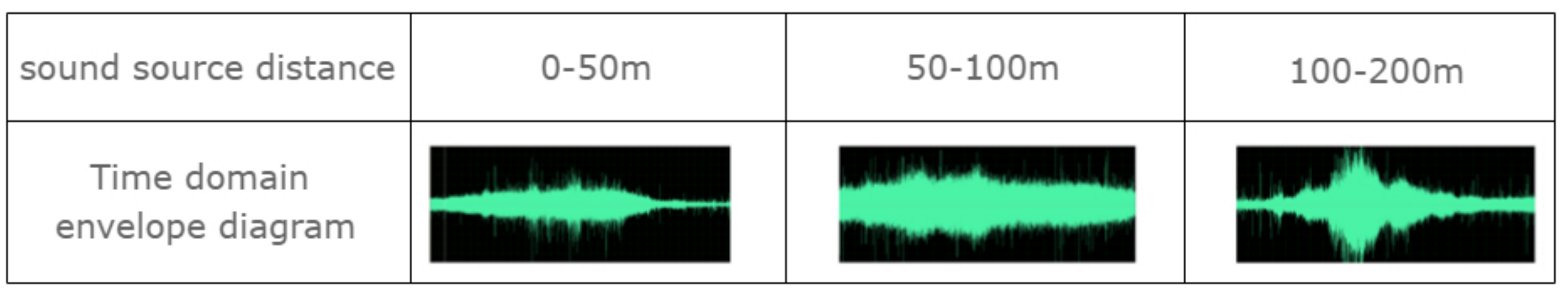}
        \centering
        \end{minipage}}
    \caption{Data analysis of sound source distance.} \label{fig:dist_analysis}
\end{figure*}

Although almost all extra information has a positive effect on UART, different types of information have different influences on the model, including distinguishing latent features, compatibility of information, or other factors. We need to dig deeper to understand what UART benefits from extra information. Next, taking ``distance of sound source'' as an example, we analyze what new knowledge does UART learn. As depicted in Fig~\ref{fig:dist_analysis}(a), data acquisition for ships with low radiated noise (Sailboat, Fishboat) is usually carried out at a close distance, and the distance tends to be relatively far for those with loud radiation noise (Dredger). The sound source distance could reflect the sound source level information and has internal relationships with the type of target. Fig~\ref{fig:dist_analysis}(b) shows that the energy of the collected sound is not inversely proportional to the distance. In actual scenarios, working conditions have a significant influence on signals. There is an obvious difference between the sound of a ship leaving port and that of a ship forwarding at full power. Due to the complexity of working conditions, the model cannot capture the distance information from superficial features (such as energy). UART has the ability to explore the connection between audio and extra information. After introducing the distance information of the sound source, it could learn new information from the audio-distance coupled data. UART can guide network training by distance information, so as to capture the latent distance information. As for other auxiliary information, many distinguishing features of information are hidden inside the audio sequence, and simple analysis cannot capture the relationship between them. The contrastive learning strategy of UART can help excavate latent features.

\subsection{Role of spectrogram encoder}

\begin{table}[ht]

	\caption{\label{tab:table3} Ablation experiments on spectrogram encoders. $w/o$ means `without', while $w/$means `with'. The experiment is carried out on Shipsear. All classification accuracy is measured in 30 seconds. We take the average accuracy on 4 folds as the metric. The audio encoder and text encoder keep unchanged.}
	\centering
		\scalebox{0.75}{\begin{tabular}{lccc}
            \hline
			\multicolumn{1}{l}{Auxiliary information} & w/o $Encoder_S$& w/ $Encoder_S$ & Benefits(+/-) \\
			\hline

			- & 86.71 & 84.26 & -2.45 \\
			Distance & 84.64  & 90.12& +5.48 \\
			Depth & 85.18 & 89.62& +4.44\\
			Local & 86.90  & 89.65& +2.75\\
            Distance,Depth &85.55 &90.69 &+5.14 \\
            Distance,Local &85.31 &90.27 &+4.96 \\
            Depth,Local &87.49 &88.74 & +1.25\\
            Distance,Depth,Local &86.07 &89.51 & +3.44\\
			\hline
         \\
		\end{tabular}}
\end{table}

We observe that the spectrogram encoder benefits contrastive learning. Generated by analyzing the audio signal, the spectrogram does not bring additional information but enables the UART encoders to learn audio from a different perspective. From Table~\ref{tab:table3}, we find the spectrogram encoder brings a 5.48\% improvement in accuracy at the most. The role of the spectrogram encoder could be viewed as an alternative data enhancement. In addition, a new encoder also enriches the composition of the loss function, the loss function is expanded from two terms to six terms. According to the results, this method does make the model more robust and reduces the probability of the model falling into the local optimum.

In the absence of auxiliary information, We note that the spectrogram encoder has negative effects. The spectrogram encoder complicates the model, while the inputs contain insufficient information. It results in the over-fitting phenomenon. In addition, with enough auxiliary information, we find that the spectrogram encoder results in smaller accuracy fluctuations. It proves that the spectrogram encoder makes the training process more stable.

\subsection{Performance on incomplete annotation set}
In practice, additional auxiliary annotations are not always readily available, the absence of some annotations is common. We conduct experiments on the incomplete annotation set. In Shipsear, the annotation of ``Wind Speed" is incomplete, more than 15 percent of the data lack related annotations. We add the ``wind speed'' into the templates, and compare the influence of the incomplete annotation set. 

As shown in Table~\ref{tab:table4}, we find that the lack of complete information provides little performance benefit to traditional methods. Incomplete information can even degrade the accuracy. But for UART, incomplete annotations could still bring appreciable improvements. It benefits from the flexibility of natural language. For the samples with missing labels, we directly delete the corresponding sentences (e.g., delete "the wind speed is {}" from the template). After deletion, the text input is still a complete and informative sentence. From the experimental results, UART is much less affected than traditional methods in the case of incomplete information. With the introduction of wind speed, the accuracy rate of UART reaches 92.74\% on Shipsear, which is the best accuracy we have ever achieved on Shipsear.

\begin{table}[ht]
    \centering
	\caption{\label{tab:table4} Performance of incomplete annotation set on Shipsear. ``wind speed'' is an incomplete annotation set,  some data (about 15\%) miss annotations. All classification accuracy is measured in 30 seconds. We take the average accuracy on 4 folds as the metric.}
		\scalebox{0.8}{\begin{tabular}{lccc}
            \hline
			Auxiliary information  &Multi-label& Multi-task & UART\\
			\hline
			- &86.15&86.15&84.26\\
			Wind Speed  &86.05&86.94&88.13\\
			Benefits(+/-) & -0.10 & +0.79 & +3.87\\
			\hline
			Distance,Depth,Local &87.87&85.05&89.51\\
			Distance,Depth,Local,Wind &88.74&83.93&\textbf{92.74}\\
            Benefits(+/-) & +0.87 & -1.12 & +3.23\\
			\hline
         \\
		\end{tabular}}
\end{table}

\subsection{Few-shot performance as a pre-training model}

\begin{table*}[ht]
	\caption{\label{tab:table5} Performance on DeepShip (cargo, passenger ship, tanker, and tug) with limited data. All classification accuracy is measured in 30 seconds. We take the average accuracy on 4 folds as the metric. 1\%, 10\%, ..., 100\% means the proportion of used data to total data.}
	\centering
	\scalebox{1}{
		\begin{tabular}{cccccccc}
            \hline
			\multicolumn{1}{l}{Model} & Pretrain weight & Tuning &1\%&10\%&25\%&50\%&100\%\\
			\hline

			\multicolumn{1}{l}{$Encoder_{A}$} & - & Fine-tune&44.33 & 61.01 & 66.64 & 67.67 & 73.14\\
			\multicolumn{1}{l}{$Encoder_{A}$}  &Multi-label& Fine-tune&46.67 & 66.52 & 70.28 & 72.02 & 73.01\\
			\multicolumn{1}{l}{$Encoder_{A}$}  &Multi-task& Fine-tune&32.90 & 62.12 & 68.17 & 71.21 & 72.43\\
            \multicolumn{1}{l}{$Encoder_{A}$}   &UART& Encoder-based&\textbf{55.26} & 71.62 & 74.84 & 75.88 & \textbf{76.56}\\
            \multicolumn{1}{l}{UART}  &UART& UART-based&52.67 & 68.16 & 70.79 & 73.16 & 74.91\\
            
            \hline
         \\
		\end{tabular}
	}
\end{table*}

\begin{figure*}
    \centering
    \includegraphics[width=0.75\linewidth]{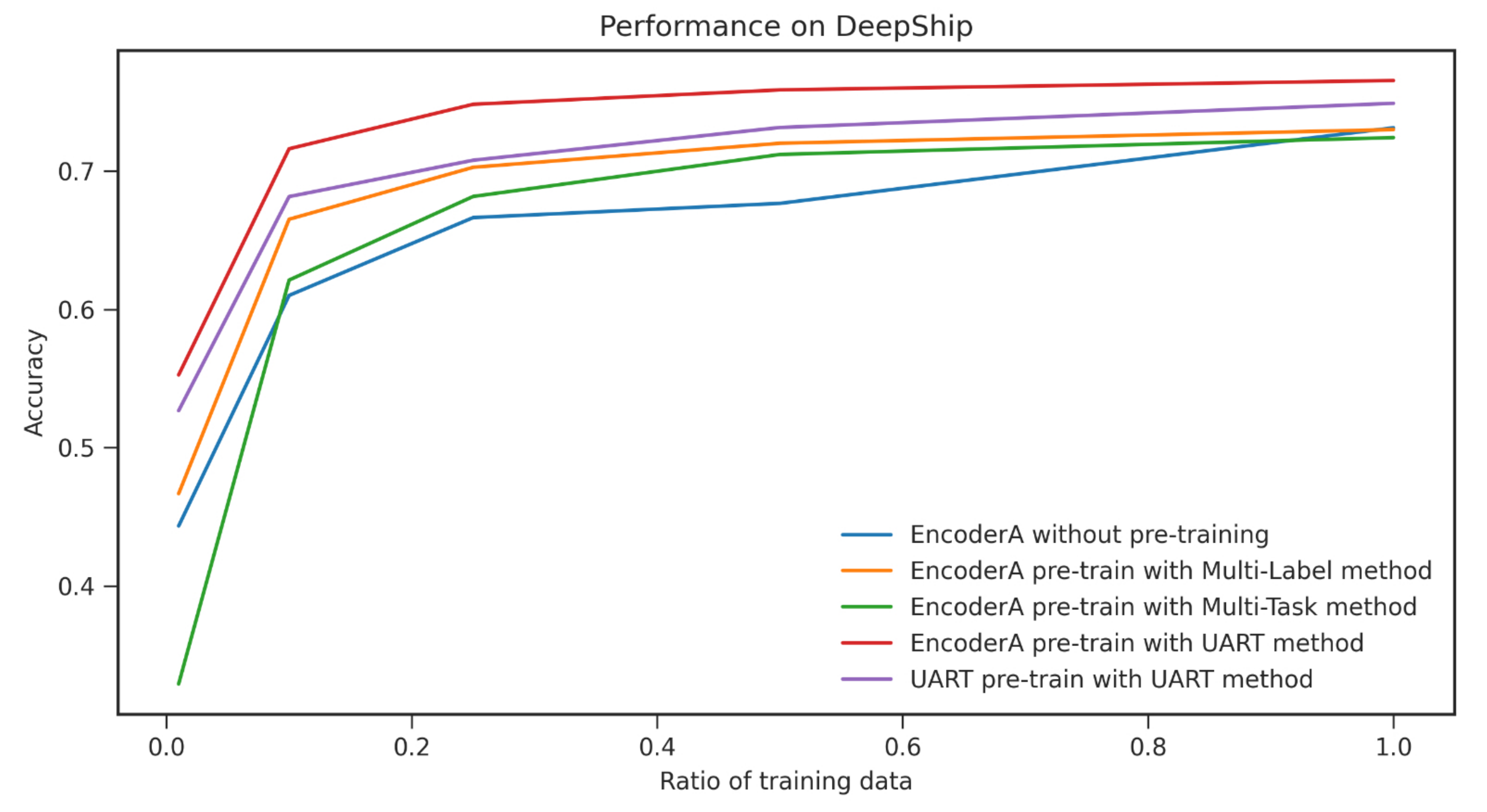}
    \caption{Performance on DeepShip with limited data. The horizontal axis represents the proportion of used data to total data.}
   \label{fig:fewshot}
\end{figure*}

For all of the above experiments, we save the model weights with good accuracy on Shipsear (including Multi-label, Multi-task, and UART). We use the pre-trained weights as the initial weights for new models. We attempt to compare the few-shot performance of the mentioned methods on the dataset without any auxiliary information. In Table~\ref{tab:table4}, we illustrate the performance of UART and baselines with limited training examples on the DeepShip dataset (there is no overlap between types of vessels in Deepship and Shipsear). We find that the audio encoder could be initialized well based on the pre-training weights of UART. The accuracy rate can reach more than 70\% with only 10\% of total training data. And UART-based pre-training weights can help the model take only 25\% of the training data to exceed the baseline approach using all training data. It shows that pre-trained UART can provide superior prior knowledge.

Besides, the few-shot performance is also amazing. UART can achieve satisfactory classification performance in the case of limited data. In addition, the model using UART's pre-training weight is still optimal when all data is used, indicating that UART can indeed endow the model with a good initialization, and make the model less prone to local optimality. 

In addition, we find that the UART-based tuning is less effective than encoder-based tuning when the training data is made up of only labels. It is because text templates seem to be redundant without auxiliary information. The conclusion fits with the results in Table.\ref{tab:table2} and the subsequent analysis. In conclusion, the best paradigm is to pre-train UART on a dataset with auxiliary information and then use the pre-trained audio encoder to tune on downstream datasets that only contain label information.


\section{Conclusion}

In this work, we explore the possibility of learning underwater acoustic representation from the descriptive natural language with contrastive learning. UART could benefit from relevant auxiliary information, thus enabling more comprehensive modeling of ship-radiated noise. By experiments on DeepShip, we demonstrate that UART could provide better prior knowledge than traditional models in practical scenarios.

There are still many points worth further study. Our current UART is only a basic version. There is still a lot of space for improvements, such as the design of text templates. Furthermore, the robustness of text input to error annotations also needs further exploration.

\section*{Acknowledgements}
This research is partially supported by the IOA Frontier Exploration Project (No. ZYTS202001), the High Tech Project (No. 31513070501) and Youth Innovation Promotion Association CAS.

\newpage

\printcredits

\bibliographystyle{cas-model2-names}

\bibliography{cas-refs-uart}





\end{document}